\documentclass[10pt,prl,aps,twocolumn,superscriptaddress,floatfix,showpacs]{revtex4}

\pdfoutput=1

\usepackage{graphicx}
\usepackage{amssymb}
\usepackage{epstopdf}
\usepackage{bm}
\usepackage{bbold}

\begin{document}

\title{Constructing gapless spin liquid state for the spin-$1/2$ $J_1-J_2$ Heisenberg model on a square lattice}

\author{Ling Wang}
\affiliation{Vienna Center for Quantum Science and Technology, Faculty of Physics, University of Vienna, Boltzmanngasse 5, 1090 Vienna, Austria}
\affiliation{Institute for Quantum Information and Matter, California Institute of Technology, Pasadena, California 91125, USA}
\author{Didier Poilblanc}
\affiliation{Laboratoire de Physique Th\'eorique, C.N.R.S. and Universit\'e de Toulouse, 31062 Toulouse, France}
\author{Zheng-Cheng Gu}
\affiliation{Institute for Quantum Information and Matter, California Institute of Technology,  Pasadena, California 91125, USA}
\author{Xiao-Gang Wen}
\affiliation{Perimeter Institute for Theoretical Physics, 31 Caroline St N,
Waterloo, ON N2L 2Y5, Canada}
\affiliation{Department of Physics, Massachusetts Institute of Technology,
Cambridge, Massachusetts 02139, USA}
\affiliation{Institute for Advanced Study, Tsinghua University, Beijing,
100084, P. R. China}
\author{Frank Verstraete}
\affiliation{Vienna Center for Quantum Science and Technology, Faculty of Physics,
University of Vienna, Boltzmanngasse 5, 1090 Vienna, Austria}

\date{\today}

\begin{abstract}
  We construct a class of projected entangled pair states (PEPS) which
  is exactly the resonating valence bond (RVB) wavefunctions endowed
  with both short range and long range valence bonds. With an
  energetically preferred RVB pattern, the wavefunction is simplified
  to live in a one parameter variational space. We tune this
  variational parameter to minimize the energy for the frustrated spin
  1/2 $J_1-J_2$ antiferromagnetic Heisenberg model on the square
  lattice. Taking a cylindrical geometry, we are able to construct
  four topological sectors with even or odd number of fluxes
  penetrating the cylinder and even or odd number of spinons on the
  boundary. The energy splitting in different topological sectors is
  exponentially small with the cylinder perimeter. We find a power law
  decay of the dimer correlation function on a torus, and a
  $\text{ln}L$ correction to the entanglement entropy, indicating a
  gapless spin liquid phase at the optimum parameter.
\end{abstract}
\pacs{75.10.Kt,75.10.Jm}
\maketitle

{\bf Introduction} -- Resonant valence bond (RVB) states, which were
first proposed by Anderson~\cite{RVB0} to describe a possible ground
state for the $S=1/2$ antiferromagnetic Heisenberg model on a
triangular lattice, and later to explain the possible mechanism of
high-$T_c$ cuprates~\cite{RVB,patrick}, provide us a rich tool box to
construct the so called spin liquid states. Rokhsar Kivelson (RK)
wavefunction~\cite{rokhsar}, which is an equal weight superposition of
nearest neighbor (NN) dimer coverings, is a critical spin liquid
state~\cite{henley} on square lattice; whereas a gaped $\mathbb{Z}_2$
spin liquid state on kagome and triangular
lattice~\cite{moessner,misguich}. The equal weight superposition of
the NN RVB state on square lattice was shown to be
critical~\cite{albuquerque,tang}. Several numerical
work~\cite{norbert1,didier,wildeboer,yangfan} have demonstrated that
the equal weight NN RVB states on the kagome and triangular lattices
are $\mathbb{Z}_2$ spin liquid states.

Recently numerical breakthroughs claimed a spin liquid ground state
for the Kagome Heisenberg model~\cite{kagomeSL,stefan} and the
frustrated spin 1/2 $J_1-J_2$ antiferromagnetic (AF) Heisenberg model
on the square lattice~\cite{jiang,ling}. However, these work did not
give direct access to the topological nature of the spin liquid
states, therefore, a simple variational wavefunction approach is
highly desirable.  Although the variational energy of the NN RVB state
on the kagome lattice~\cite{didier,yangfan} is still higher than the
energy obtained via the density matrix renormalization group (DMRG)
method~\cite{kagomeSL}, the topological nature is well understood
within the formalism of the projected entangled pair states
(PEPSs)~\cite{didier}.  On the other hand, from a projective
wavefunction~\cite{liang} approach supplemented by a projective
symmetry group (PSG) analysis all possible spin-liquid states on the
triangular~\cite{fa1}, Kagome~\cite{fa1,ran,lesik1,lesik2} and
Honeycomb~\cite{fa2} lattices have been obtained and classified but,
for all lattices, the energetically favorable states are believed to
involve longer range RVB.  As a result, it is natural to think that a
general RVB state within the PEPS formalism is a more practical
variational wavefunction, where one can gain simultaneously an
optimized energy {\it and} a comprehensive picture of the topological
nature.

In this paper, we introduce a general RVB state written as a $D=3$
PEPS, different from Ref.~\cite{norbert1,didier}, i.e. it includes
valence bonds of all length (although with a bond amplitude decaying
exponentially fast with the bond length). With a properly chosen
singlet sign convention that meets all lattice symmetries on the
square lattice, we minimize the energy of the spin 1/2 $J_1-J_2$ AF
Heisenberg model at $J_2=0.5J_1$ against a single variational
parameter $c$ governing the decay amplitude of the long range valence
bonds.  The idea is therefore to introduce a simple yet competing
wavefunction that enables us to fully understand the topological
properties of the ground state of the frustrated magnets.

\begin{figure}
\begin{center}
\includegraphics[width=8cm]{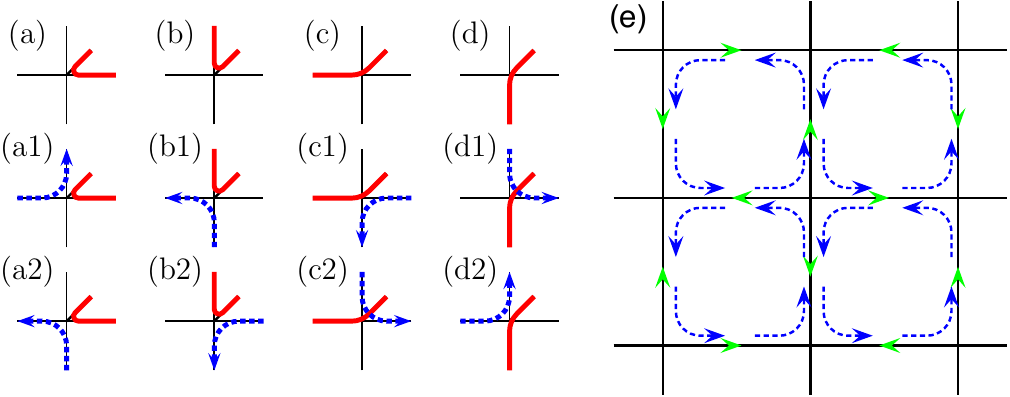}
\caption{ (a-d) Red thick lines denote mappings of a virtual spin $1/2$ to a
  physical spin $1/2$, the blue dash lines represent the
  singlet pairings of two virtual spin-$1/2$s, and the black thin lines
  represent the virtual spin 0 states. (e) The local sign convention for the bond singlets (in green arrows) and the corner singlets (in blue arrows). }
\label{vertices}
\end{center}
\vskip -0.5cm
\end{figure}

{\bf RVB states in PEPS formalism} -- The equal weight superposition
of the NN RVB states can be constructed using a PEPS with bond dimension
$D=3$ as following: each physical site has 4 virtual spins attached,
each of which spans a virtual dimension of spin $1/2\oplus 0$. From
the bond point of view, every pair of the NN virtual spins is
projected to a block diagonal virtual spin singlet state:
\begin{equation}
|\mathcal{S}\rangle = |01\rangle-|10\rangle +|22\rangle,
\end{equation}
here the virtual indices ``0,1'' span the subspace of spin $1/2$ and
virtual index ``2'' spans the subspace of spin 0. At the physical
site, a projector enforces one of the virtual spins with its spin
$1/2$ subspace to be mapped to the physical spin $1/2$ state and the
rest of virtual spins to stay in the spin 0 subspace, {\it i.e.}  the
``2'' state,
\begin{equation}
\label{projector}
\mathcal{P}_1=\sum_{k=1}^4(|\uparrow\rangle\langle 0|_k+|\downarrow\rangle\langle 1|_k)\otimes \langle 222|_{/k},
\end{equation}
here subscript ``$/k$'' stands for all except $k$. This PEPS, by
contracting the virtual index of each $\mathcal{S}$ at the bond and
each $\mathcal{P}_1$ at the vertex, represents exactly the equal
weight NN RVB states.

To allow long distance singlet pairings, we need spins to teleportate:
enforcing a singlet between site $i$ and $j$ that are already paired in
singlets $(s_1,i)$ and $(s_2,j)$ will generate a singlet pair
$(s_1,s_2)$. The following projector realizes spin teleportation
without increasing the bond dimension,
\begin{equation}
\nonumber
  \mathcal{P}_2=\sum_{i\neq j\neq k\neq l}(|\uparrow\rangle\langle 0|_i+|\downarrow\rangle\langle 1|_i)\otimes\langle 2|_j\otimes\langle\epsilon|_{kl},
\end{equation}
here $|\epsilon\rangle_{kl}\equiv |01\rangle_{kl}-|10\rangle_{kl}$,
and it forces spins connected via this site by bonds $k$ and $l$ into
a singlet. A general RVB wavefunction is a parameter $c$ weighted
combination of projectors
$\mathcal{P}\equiv\mathcal{P}_1+c\mathcal{P}_2$ at each vertex
$V\equiv\{v\}$ traced out with the bond singlets $\mathcal{S}$ at
each bond $B\equiv\{b\}$,
\begin{equation}
|\Psi\rangle_{\text{RVB}}= \prod_V\mathcal{P} \prod_ B|\mathcal{S}\rangle,
\label{non-vison}
\end{equation}

{\bf The sign convention and symmetries} -- Fig.~\ref{vertices}(a-d)
enumerates 4 possible $\mathcal{P}_1$ projectors and 8 $\mathcal{P}_2$
projectors at each vertex. The bond singlet $\mathcal{S}$ is chosen
such that NN singlets point from sublattice A to B; the corner
singlets, which plays the role of singlet teleportation, are oriented
counter clockwise and preserve all lattice symmetries.
The sign convention is demonstrated in Fig.~\ref{vertices}(e).

\begin{figure}
\begin{center}
\includegraphics[width=8cm]{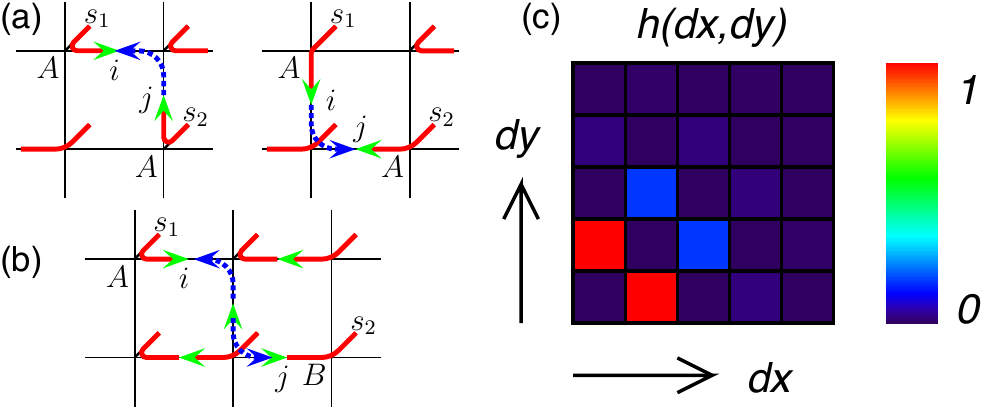}
\caption{(a) The NNN singlet $(s_1,s_2)$ pairs through 2 bond singlets
  and 1 corner singlet via 2 paths, which cancel each other. (b) The
  next allowed AB sublattice pairing $(s_1,s_2)$ through 3 bond
  singlets and 2 corner singlets. (c) The weight distribution in the
  Liang-Doucot-Anderson picture with a linear color scale for a
  $8\times 8$ torus at $c=0.35$. $h(dx,dy)$ is the weight of a singlet
  at separation $(dx,dy)$, where $h(1,0)=1$. The plotted color scale
  takes the square root of $h(dx,dy)$ to magnify the weight of the
  long range singlet.}
\label{sign}
\end{center}
\vskip -0.5cm
\end{figure}

The NNN singlet arises through two bonds singlets and one corner
singlet, as in Fig.~\ref{sign}(a). However the weight of a diagonal
singlet is comprised of two shortest paths of equal magnitude but
opposite sign, thus the net weight of the diagonal singlet is
zero. The only shortest path to build the next range AB sublattice
singlet is shown in Fig.~\ref{sign}(b), and it consists of three bond
singlets and two corners. The sign of the next range AB singlet is
pointing from sublattice A to B. In general, no AA pairings survive
(see Appendix) and all AB pairings point from sublattice A to B. To
verify this result, we implement a Monte Carlo (MC) sampling of the
singlet distribution of (\ref{non-vison}) and calculate the weight
$h(dx,dy)$ defined in Ref.~\cite{liang} as a function of
separation. The result is presented in Fig.~\ref{sign}(c) and is
consistent with the above analysis.

{\bf String picture} -- Since the RVB bonds only connect sites on
different sublattices, we can view such a RVB state as a liquid state
of oriented strings. Indeed, choosing a reference $\text{VB}_0$
configuration, any $\text{VB}$ configuration can be viewed as a closed
oriented string configuration: the RVB bonds in the $\text{VB}$
configuration are regarded as a piece of the closed string pointing
from the A to the B sublattice, while the RVB bonds in the reference
$\text{VB}_0$ configuration are regarded as the complementary piece
pointing from the B to the A sublattice.

If such a superposition of closed orientable string states indeed
represent a \emph{liquid} state of closed strings, then the
entanglement entropy for a such a state in a region $A$ has the form
$S_A= a L_A - \frac 12 \text{ln} L_A+b$, where $L_A$ is the length of
the perimeter of region $A$. To understand such a result, we view a
string as a flux line and the close-string condition implies that the
flux is conserved.  Therefore, the total flux going through the
perimeter of $A$ is zero. If we had ignored the flux conservation and
assumed that the flux could fluctuate freely and independently, then
the entanglement entropy would have an exact area-law form $S_A=
aL_A$ and the typical amount of flux through the perimeter would be
proportional to $\sqrt{L_A}$. So if we restrict the amount of flux through
the perimeter to be zero, the entanglement entropy will be $S_A= a L_A
- \text{ln} \sqrt{L_A}+b= a L_A - \frac 12 \text{ln} L_A+b$.  The
ln$L_A$ dependence in the entanglement entropy implies that the liquid
of orientable closed strings must be gapless. This property is
confirmed by our numerical calculation.

\begin{figure}
\begin{center}
\includegraphics[width=8.7cm]{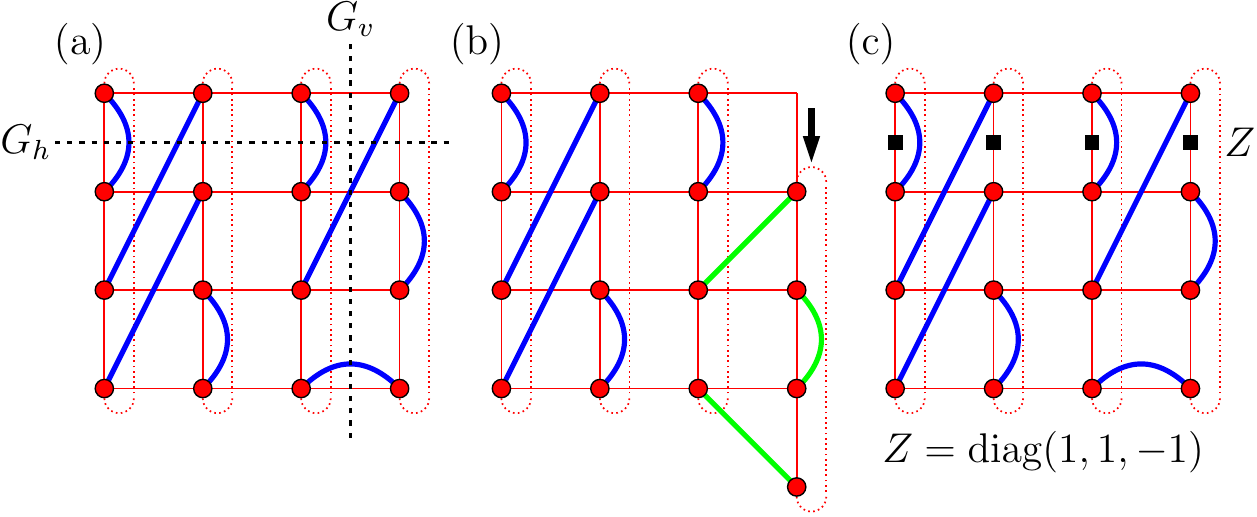}
\caption{(a) The parity quantum number $G_{v}$ ($G_h$) is defined by
  counting the number of singlets (blue lines) modulo 2 crossed by a
  vertical loop (horizontal line) in dashed black line. (b) By
  cyclically permuting all spins in a loop winding around the
  cylinder, a $G_h=1$ configuration in (a) is transformed into a
  $G_h=-1$ configuration in (b).  Singlets in green lines are affected
  by the permutation operator. (c) An odd-flux state is defined by
  inserting operator $Z$ on all vertical bonds crossed by a horizontal
  line to (\ref{non-vison}).}
\label{permutation}
\end{center}
\vskip -0.5cm
\end{figure}

{\bf Variational ground state energies at $J_2=0.5J_1$} -- We consider
the general RVB wavefunction on a cylinder with finite cylindrical
circumference $N_v$ and infinite horizontal length $N_h=\infty$. The
physical properties are determined by the eigenvector with the largest
eigenvalue of the transfer matrix. Let us introduce a horizontal
(vertical) parity number $G_h$ ($G_v$) which is defined by counting
the number of singlets modulo 2 that cross a horizontal (vertical)
line joining the two boundaries of the cylinder (going around the
cylinder). The two states with $G_h=\pm 1$ ($+$ is even and $-$ is
odd) are orthogonal to each other in the thermodynamic limit, and can
be transformed from one to another by a cyclic spin permutation
$\Pi_{\bigcirc}$ operator winding around the cylinder as illustrated
in Fig.~\ref{permutation}(a,b).

The $G_h=\pm 1$ states are not the minimally entangled states
(MESs)~\cite{vishwanath}. However their superpositions,
\begin{equation}
|\Psi(\pm)\rangle\equiv|\Psi\rangle_{G_h=1}\pm |\Psi\rangle_{G_h=-1},
\label{Eq:pm}
\end{equation}
with a relative $\pm$ sign are. A good reason for it is that these
states (\ref{Eq:pm}) can be written as simple PEPSs: $|\Psi(+)\rangle$
state is the state corresponding to (\ref{non-vison}), and
$|\Psi(-)\rangle$ state is obtained by inserting a ``vison'' line to
the PEPS for state $|\Psi(+)\rangle$ (by putting
$Z=\text{diag}(1,1,-1)$ operators on the vertical bonds crossed by a
horizontal line), as in Fig.~\ref{permutation}(c). These MESs are
referred to as the even-flux ($|\Psi(+)\rangle$) states and the
odd-flux ($|\Psi(-)\rangle$) states which stands for even and odd
number of flux penetrating the cylinder. We will show next that these
four states $|\Psi(\pm 1)\rangle_{e/o}$ are (bulk) ground states of
the gapless spin liquid state.

\begin{figure}
\begin{center}
\includegraphics[width=8.7cm]{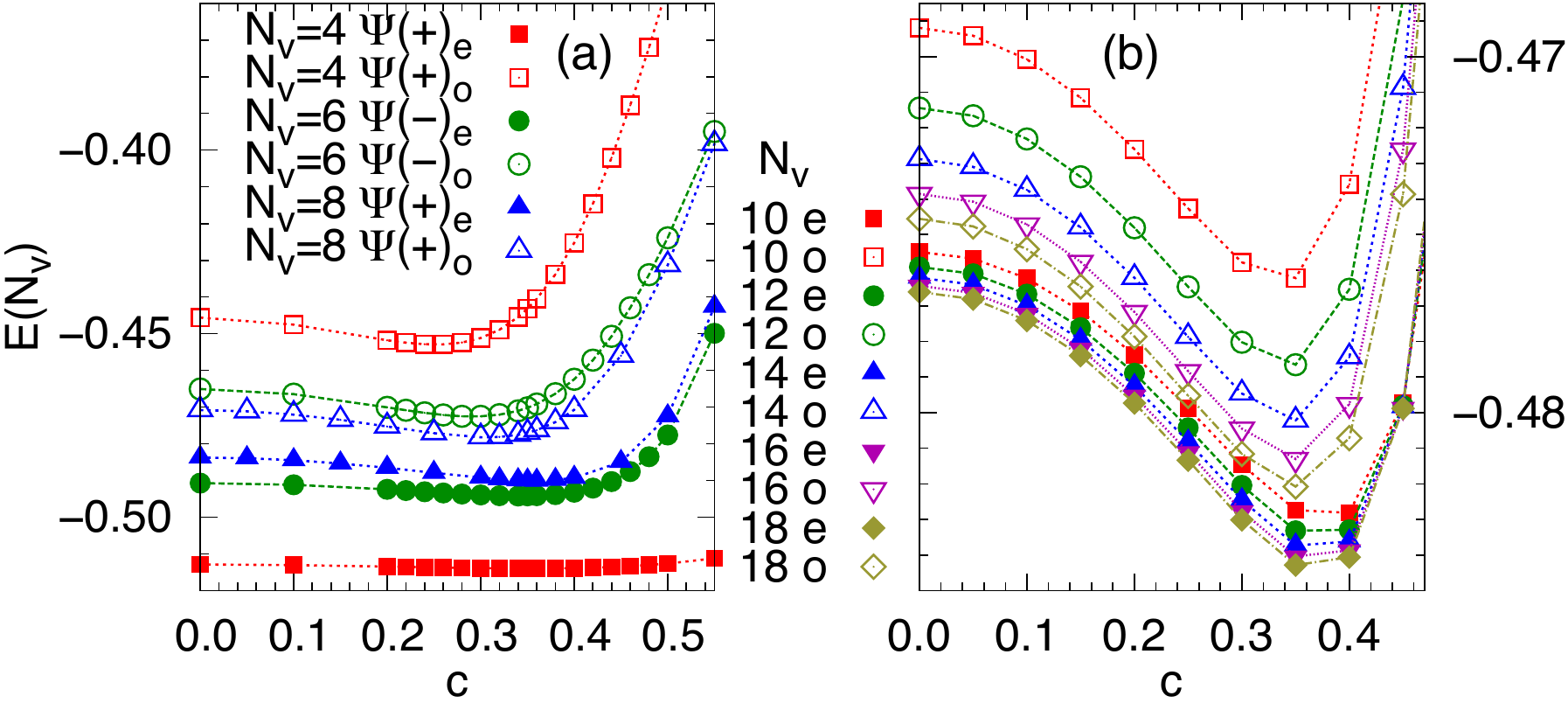}
\caption{Ground state energies (per site) computed for the $J_1-J_2$ AF
  Heisenberg model at $J_2=0.5J_1$, as a function of parameter
  $c$ for the $e/o$ topological sectors of (a) a cylinder ($N_v=4,6,8$)
  with even-flux ($N_v/2$ even) or odd-flux ($N_v/2$ odd), and (b)
  a strip geometry with $N_v=10,\cdots,18$.
  For both cases, energy minimizes at $c=0.35(1)$.}
\label{enr1}
\end{center}
\vskip -0.5cm
\end{figure}

The variational energies of $|\Psi(\pm)\rangle_{e/o}$ on cylinders
with finite perimeter $N_v=4,6,8$ ($N_v$ must be even, otherwise the
system dimerizes) are computed {\it exactly} via the transfer matrix
method and shown in Fig.~\ref{enr1}(a) (an even or odd number of flux
is chosen to provide the lowest energy) as a function of the
variational parameter $c$. The best variational energy for the spin
$1/2$ $J_1-J_2$ AF Heisenberg model at $J_2=0.5J_1$ is $c=0.35(1)$
with $N_v=4,6,8$. To access larger system size, we study a
complementary geometry where the cylinders are cut open, with the top
and bottom vertical virtual spins set to ``2''s. We call them the
finite ($N_v$) width strips. For a contractible geometry as strips,
the flux parity is no longer meaningful, but the boundary parity
quantum number $G_v=e(o)$ still holds. We simulate the leading
eigenvectors of the transfer matrix of the strips by the matrix
product states (MPS) with the same quantum number $G_v$ in both bra
and ket. The ground state energies as a function of the variational
parameter $c$ for both sectors ($|\Psi\rangle_{e/o}$) of the strips
with $N_v=10,\cdots,18$ are presented in Fig.~\ref{enr1}(b). The best
variational energy for $J_2=0.5J_1$ is at $c=0.35$, which is in good
consistency with the case of finite perimeter cylinders.

\begin{figure}
\begin{center}
\includegraphics[width=8.7cm]{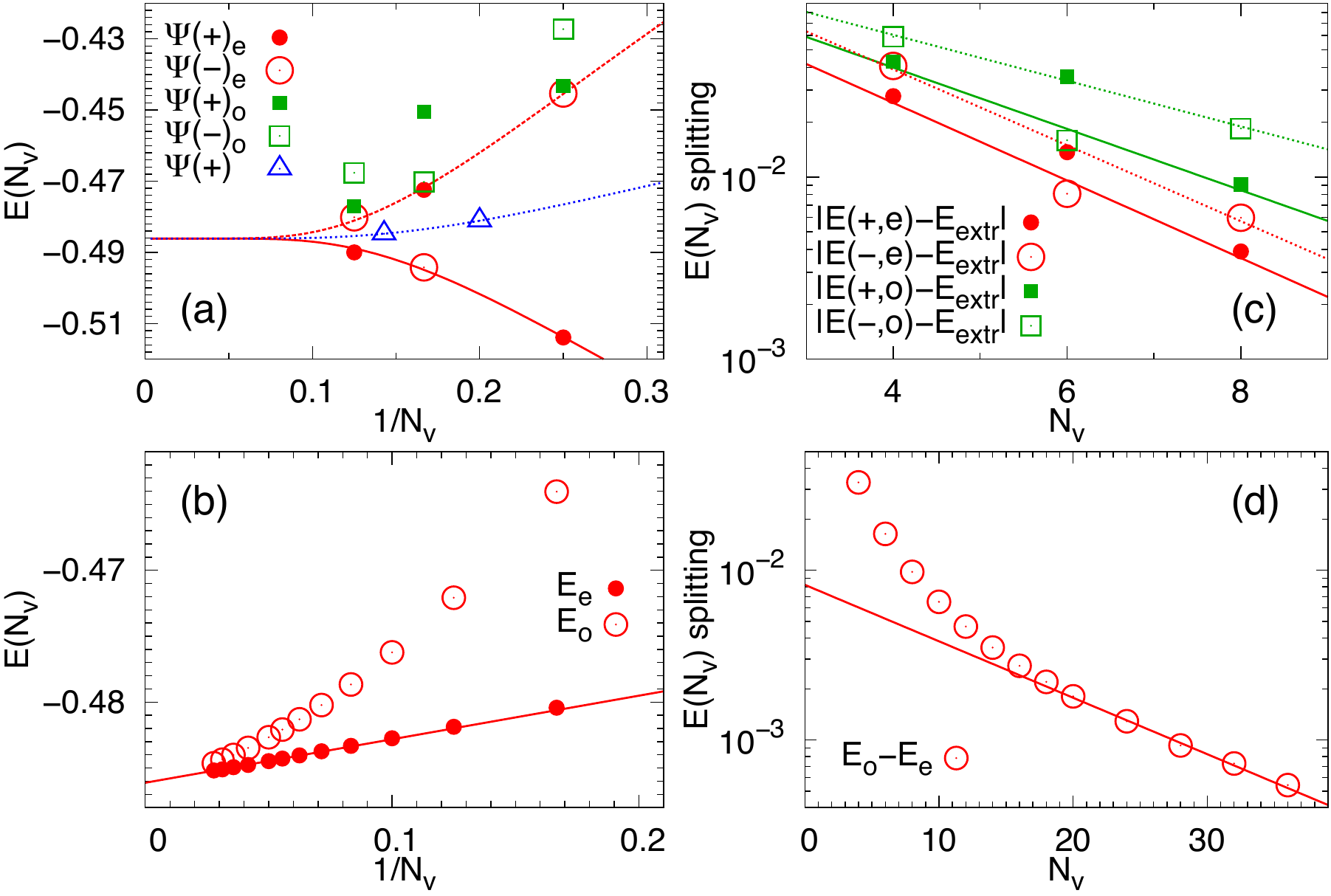}
\caption{ Ground state energies at $c=0.35$ as a function of $1/N_v$
  for all topological sectors in (a) cylinders and (b) strips. The
  extrapolated ground state energy from the even sector of strips is
  $E_{\infty}=-0.48612(2)$. The ground state energy splitting between
  the lowest energy sector and other sectors as a function of $N_v$
  for (c) cylinders and (d) strips. The splitting vanishes
  exponentially with size $N_v$.}
\label{enr2}
\end{center}
\vskip -0.5cm
\end{figure}

The variational energies of the even and odd sectors at the optimum
parameter $c=0.35$ as a function of inverse width $1/N_v$ are shown in
Fig.~\ref{enr2}(a,b), with cylinders of size $N_v=4,6,8$ and strips up
to $N_v=36$. A linear regression is applied to the even sector of the
strips and a thermodynamic limit of $E_{\infty}=-0.48612(2)$ is
obtained. This energy is competing on the third decimal digit to the
best variational estimate of $E_{\infty}=-0.4943(7)$ with a $D=9$
PEPS~\cite{ling}, let alone the fact that here we vary only one
variational parameter in a $D=3$ PEPS. A conjecture about the ground
state energies of the gapless and gaped spin liquid states is that the
energy splittings between different topological sectors become
exponentially small with the system size. This conjecture is verified
in Fig.~\ref{enr2}(c,d) presenting on semi-log scales the energy
difference between all sectors and $E_{\infty}$ for cylinders and
between the two existing energy sectors for strips.

\begin{figure}
\begin{center}
\includegraphics[width=8.7cm]{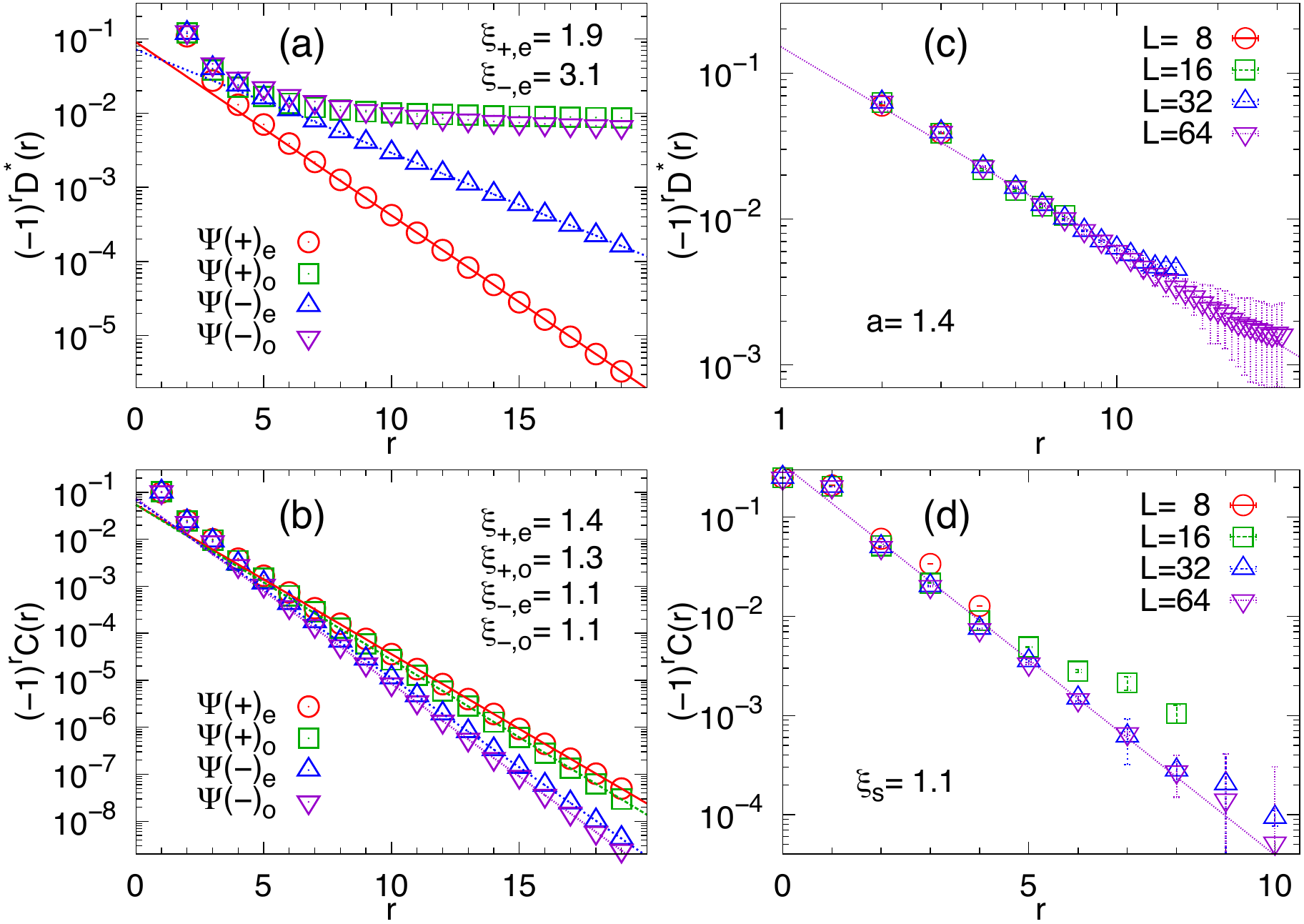}
\caption{The dimer (a) and the spin (b) correlation function for all
  topological sectors on a $N_v=8$ cylinder at $c=0.35$. The dimer (c)
  and the spin (d) correlation function on a torus with $L=8,16,32,64$
  at $c=0.35$. Note that different scales, log-log in (c) and semi-log
  otherwise, are used. }
\label{ddc1}
\end{center}
\vskip -0.5cm
\end{figure}

\begin{figure}
\begin{center}
\includegraphics[width=5cm]{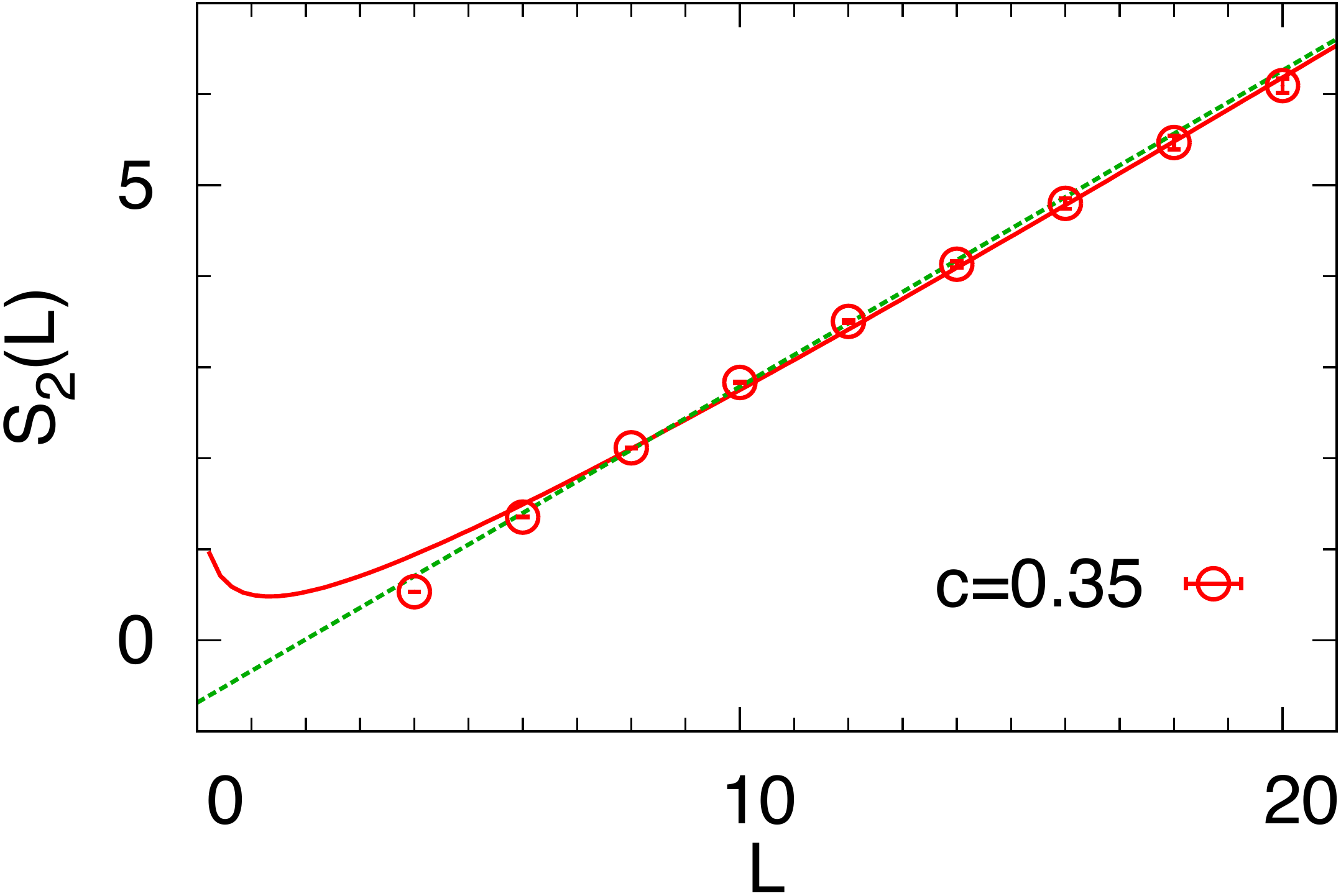}
\caption{Renyi entanglement entropy $S_2(L)$ for an area $L\times L$
  on a torus of size $2L\times L$ with $L=4,\cdots,20$. The fitted red
  line is of form $S_2(L)=a_1L-\frac 1 2 \ln L+b_1$ for $L\in [6:18]$; the
  fitted green line is $S_2(L)=a_2L+b_2$ for $L\in [6:14]$ and
  $b_2=-0.68(1)$.}
\label{entropy}
\end{center}
\vskip -0.5cm
\end{figure}

{\bf Correlation functions and entanglement entropy} -- We define the
spin and dimer correlation functions as the ground state expectation
values $ C(r)=\big< \mathbf{S_0}\cdot\mathbf{S}_r\big> $ and $
D^*(r)=\big<
(\mathbf{S_0}\cdot\mathbf{S}_1)(\mathbf{S_r}\cdot\mathbf{S}_{r+1})-(\mathbf{S_0}\cdot\mathbf{S}_1)
(\mathbf{S_{r-1}}\cdot\mathbf{S}_{r}) \big> $.  Fig.~\ref{ddc1}(a)
plots the dimer correlation functions on a cylinder with $N_v=8$ for
all topological sectors at the optimal parameter $c=0.35$. The odd
sectors have very slowly decaying dimer correlations due to an odd
number of spinons sitting on the boundaries, thus the system
effectively becomes an odd-width cylinder and the Majumdar-Ghosh kind
of degeneracy emerges. We can eliminate the boundary effect by setting
the system on a torus and carrying a variational MC simulation for
PEPSs~\cite{lingmc}. We found the dimer correlation function exhibits
a power law decay $D^*(r)\sim\frac{(-1)^r}{r^a}$ with $a=1.4$, as
shown in Fig.~\ref{ddc1}(c). In contrast, the decay of the spin
correlation function for all sectors on a cylinder or on a torus
remains exponential with a correlation length $\xi_s\approx 1.1$, as
evidenced in Fig.~\ref{ddc1}(b,d). Fig.~\ref{entropy} shows the Renyi
entropy $S_2(L)$ of an area $L \times L$ on a $2L\times L$ torus for
size $L=4,6,\cdots,20$. The fitted line of $aL-\frac 1 2\text{ln}L+b$
reflects the $\text{ln}L$ correction from the oriented string
picture. The simulation is done via MC sampling of the RVB
configuration~\cite{roger}. Finally, we also would like to point out
that the logarithmic correction is very hard to be detected on small
system size. If we fit $S_2$ with a form $aL+b$ on small system size,
we find that the constant $b=-0.68(1)$, which is very close to
$-\text{ln}2$. Such an observation implies that the observed
$-\text{ln}2$ constant in DMRG calculation~\cite{jiang} is
insufficient to rule out the possibility of gapless spin liquid ground
state for $J_1-J_2$ Heisenberg model on square lattice.

{\bf Conclusion and outlook} -- We constructed a class of projected
entangled pair states which exactly represent general RVB
wavefunctions with all bond length contributions. Upon choosing an
energetically preferred RVB pattern, we are able to build a
one-parameter manifold of variational RVB $D=3$ PEPSs which preserve
all lattice symmetries.  Minimization of the variational energy for
the frustrated spin $1/2$ $J_1-J_2$ Antiferromagnetic (AF) Heisenberg
model on the square lattice yields, at $J_2=0.5J_1$, an energy
$E_{\infty}=-0.48612(2)$ per site in the thermodynamic limit. In the
case of a cylinder geometry, four orthogonal topological states were
identified, namely the even-flux and odd-flux states with even and odd
number of spinons on the boundary. We found the dimer correlation
function decays algebraically while the spin correlation function
still decays exponentially.  The entanglement entropy scaling reveals
$\text{ln}L$ correction to the area law. Both features point towards
the gapless spin liquid nature of our constructed RVB wavefunction.

Previous valence bond MC simulations have proposed wavefunctions which
violate the Marshall's sign rule by a single negative pairing
magnitude $h(2,1)$~\cite{jie,kevin}, however our PEPS wavefunction
constructed to meet a negative $h(2,1)$ condition does not gain an
optimized energy except on a very small $4\times 4$ torus.

The PEPSs construction of the general RVB states can be applied to
other bipartite and non-bipartite lattices, where the Schwinger boson
spin liquid states under the projective symmetry group (PSG) analysis
have been found~\cite{fa1,ran,lesik1,lesik2,fa2,tao}, but for which
thermodynamic energies and correlation functions are still unknown due
to a negative sign problem in the valence bond MC simulations. Within
the PEPS formalism all of these can be easily studied. Our PEPS
construction of the RVB states can be further generalized to
accommodate more complicated pairing pattern which can improve further
the ground state energy although possibly requiring a larger bond
dimension.

{\bf Acknowledgment} -- We would also like to thank N. Schuch,
I. Cirac, D. Perez-Garcia and O. Motrunich for stimulating
discussions. This project is supported by the EU Strep project
QUEVADIS, the ERC grant QUERG, the FWF SFB grants FoQuS and ViCoM, and
the NQPTP ANR-0406-01 grant (French Research Council).
XGW is supported by NSF Grant No. DMR-1005541, NSFC 11074140, and
NSFC 11274192.  Research at Perimeter Institute is supported by the
Government of Canada through Industry Canada and by the Province of Ontario
through the Ministry of Research.
The
computational results presented have been achieved using the Vienna
Scientific Cluster (VSC) and the CALMIP Hyperion Cluster (Toulouse).

\begin{widetext}
\section*{Supplementary material}
\subsection*{$|\Psi\rangle_{\text{RVB}}$ has no $AA$ ($BB$) pairing}
In the main text, we have illustrated why the leading order contribution of NNN pairing vanishes in $|\Psi\rangle_{\text{RVB}}$. Here we would like to show that, indeed, all the $AA$ ($BB$) pairing vanish in $|\Psi\rangle_{\text{RVB}}$ exactly up to any order in $c$. As explained in the text, the pairing amplitude between two sites $i$ and $j$ is given by the sum of
all teleportation paths that connect $i$ and $j$. As seen in Fig. \ref{pairing}, the solid red line a is a typical teleportation path that contributes to the pairing $(ij)$. According to the local rule of the teleportation path (each path must turn at a vertex), it is not hard to see that any teleportation path contributing to the $AA$ ($BB$) pairing must go through an odd number of corners and an even number of links. On the other hand, for a given teleportation path a, we can always find a dual teleportation path b which goes though the same number of corners and links. For any pair of dual a and b paths, the sign contribution from all the links (corners) are the same (opposite). Therefore, the total contribution from the pair of teleportation paths a and b always vanishes. Thus, we have proved that $|\Psi\rangle_{\text{RVB}}$ has no $AA$ ($BB$) pairing.
\begin{figure}
\begin{center}
\includegraphics[width=4cm]{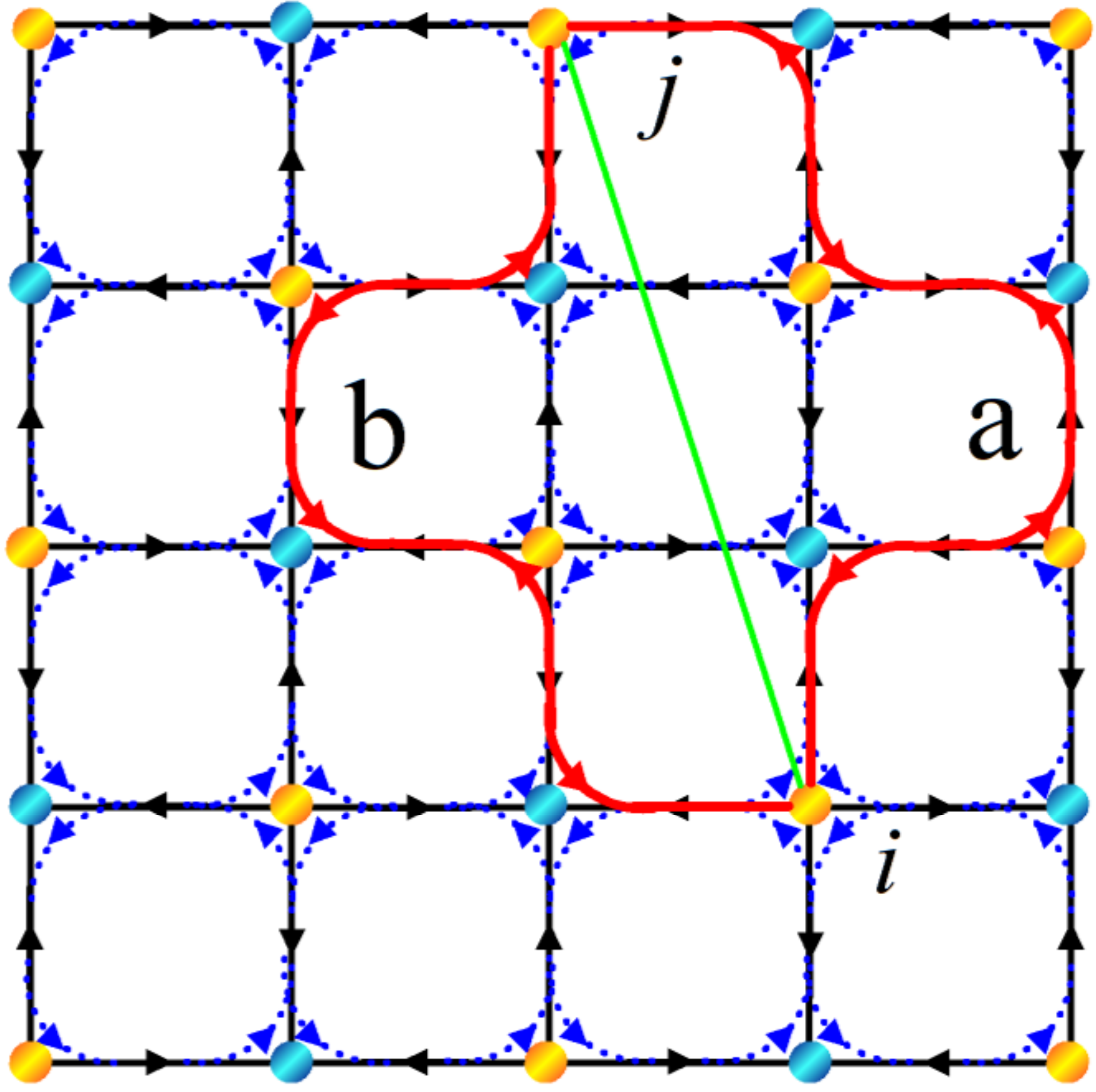}
\caption{(Color online) For any teleportation path contributing to the $AA$ ($BB$) pairing, there always exists a dual teleportation path with opposite sign. Therefore
the $AA$ ($BB$) pairing amplitudes are exactly zero in $|\Psi\rangle_{\text{RVB}}$.}
\label{pairing}
\end{center}
\end{figure}

\subsection*{Finite size extrapolation of the optimum parameter $c$}
We provide here a more accurate determination of the optimum parameter $c$.
For this purpose, we use a quadratic function to fit the even sector ($|\Psi\rangle_e$)
energy of a finite width ($N_v$) strip as a function of parameter $c$ (around the minimum),
and we extract the optimum parameter $c_{\text{opt}}(N_v)$ and the
minimum energy $E_{\text{min}}(N_v)$ for $N_v=10,\cdots,24$. The
results are presented in Fig.~\ref{para}(a). Taking the optimum
parameter $c_{\text{opt}}(N_v)$ and extrapolating it to the thermodynamic
limit as a function of inverse width $1/N_v$, we find
$c_{\text{opt}}(\infty)=0.356(1)$ (see Fig.~\ref{para}(b)). Again,
taking the minimum finite width energy $E_{\text{min}}(N_v)$ and
extrapolating it to the thermodynamic limit as a function of $1/N_v$, we
obtain a thermodynamic energy $E_{\infty}=-0.48620(1)$ (see
Fig.~\ref{para}(c)).

\begin{figure}
\begin{center}
\includegraphics[width=7in]{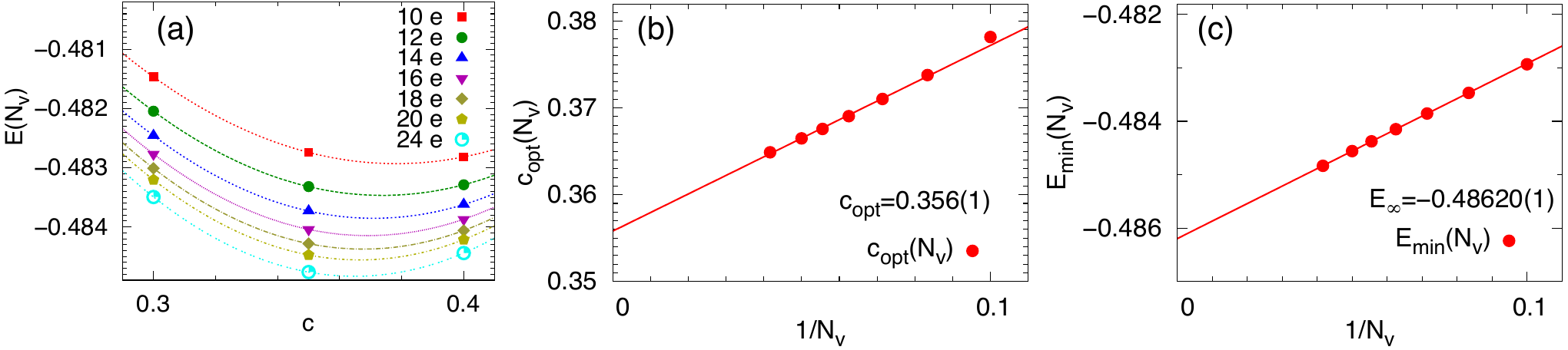}
\caption{(Color online)(a) A quadratic fit of the even sector energy $E(N_v)$ of a strip as a function of $c$ around its minimum with $N_v=10,\cdots,24$. (b) Optimum parameter $c_{\text{opt}}(N_v)$ plotted as a function of $1/N_v$. A linear regression gives $c_{\text{opt}}(\infty)=0.356(1)$. (c)  Minimum energy $E_{\text{min}}(N_v)$ plotted as a function of $1/N_v$. A linear regression gives $E(\infty)=-0.48620(1)$.}
\label{para}
\end{center}
\end{figure}

\end{widetext}

\end{document}